\documentclass[aps,prd,onecolumn,showkeys,showpacs,amssymb]{revtex4}
\usepackage{float}
\usepackage{amssymb}
\usepackage{graphicx}
\usepackage{amsmath}
\usepackage{hyperref}
\begin{document}

\newcommand{\be}{\begin{equation}}
\newcommand{\ee}{\end{equation}}

\title{Nonperturbative models of quark stars in $f(R)$ gravity}

\author{Artyom V. Astashenok$^{1}$, Salvatore Capozziello$^{2,3,4}$, Sergei D. Odintsov$^{5,6}$}

\affiliation{$^{1}$I. Kant Baltic Federal University, Institute of Physics and Technology, Nevskogo st. 14, 236041 Kaliningrad, Russia,\\
$^2$Dipartimento di Fisica, Universita' di Napoli "Federico II
and \\$^3$INFN Sez. di Napoli, Compl. Univ. di Monte S. Angelo, Ed. G., Via Cinthia,
9, I-80126, Napoli, Italy,\\
$^4$ Gran Sasso Science Institute (INFN), Viale F. Crispi, 7, I-67100. L'Aquila, Italy.\\
$^5$Instituci\`{o} Catalana de Recerca i Estudis Avan\c{c}ats (ICREA), Barcelona, Spain\\
$^6$Institut de Ciencies de l'Espai (IEEC-CSIC), Campus UAB,
Facultat de Ciencies, Torre C5-Par-2a pl, E-08193 Bellaterra,
Barcelona, Spain}

\begin{abstract}
Quark star models with realistic equation of state in
nonperturbative $f(R)$ gravity are considered. The mass-radius
relation for $f(R)=R+\alpha R^2$ model is obtained.  Considering scalar
curvature $R$ as an  independent function, one can find out, for each value
of central density,  the unique value of central curvature for which
one has solutions with the required asymptotic $R\rightarrow 0$ for
$r\rightarrow\infty$. In another words, one needs a fine-tuning for
$R$ to achieve quark  stars in $f(R)$ gravity. We consider also the analogue
description in corresponding scalar-tensor gravity. The
fine-tuning on $R$ is equivalent to the  fine-tuning on the scalar field
$\phi$ in this description. For distant observers,  the gravitational
mass of the  star increases with increasing $\alpha$ ($\alpha>0$) but the
interpretation of this fact depends on frame where we work.
Considering directly $f(R)$ gravity,  one can say that increasing of
mass occurs by the ``gravitational sphere'' outside the star with some
``effective mass''. On the other hand, in conformal scalar-tensor theory,  we
also have a dilaton sphere (or "disphere") outside the star but its
contribution to gravitational mass for distant observer is
negligible. We show that it is possible to discriminate modified
theories of gravity from General Relativity due to the gravitational
redshift of the thermal spectrum emerging from the surface of the star.

\end{abstract}



\date{\today}

\maketitle

\section{Introduction}

The accelerated expansion of the Universe remains one of the
puzzles of modern cosmology. Initially discovered by observations
of distant standard candles \cite{Perlmutter, Riess1, Riess2},  this
acceleration is confirmed by several another observations such as
microwave background radiation (CMBR) anisotropies \cite{Spergel},
cosmic shear through gravitational weak lensing surveys
\cite{Schmidt} and data on Lyman alpha forest absorption lines
\cite{McDonald}. Analysis of these observations shows that the
required cosmological dynamics cannot be obtained by models where the  universe contains only standard matter and radiation or, in some sense,  canonical scalar
fields.

One  possible solution of this puzzle is that General
Relativity should be modified. It is possible to obtain
accelerated expansion in modified gravity without assuming dark
energy  as a new material field \cite{Capozziello1, Capozziello2, Odintsov1, Turner, Olmo,
Odintsov-3, Capozziello_book, Capozziello4, Cruz}.

Another explanation considers  the existence of a non-standard
cosmic fluid with negative pressure  consisting about
70\% of the universe  energy, which is not clustered in large
scale structure. However, the nature of this dark energy fluid  is unclear. According
to the simplest hypothesis, the dark energy is nothing else but the Einstein
Cosmological Constant. Despite of some questions at fundamental
level (for example the cosmological constant problem and problems with fine tuning \cite{Weinberg}),
the $\Lambda$CDM model
\cite{Bancall}, based on dark matter and cosmological $\Lambda$ term,   gives, in principle,  a well agreement with
observational data at present epoch.

It has been shown that modified gravity also could give adequate
description of cosmological observations \cite{Bancall,Demianski, Perrotta, Hwang}. One can conclude therefore that
cosmological observations only cannot witness in favor to modified
gravity or $\Lambda$CDM model. We need new probes and testbeds at completely different scales.

Specifically,  any theory of modified gravity should be tested at
astrophysical level also.  One can hope that strong field regimes
of relativistic stars could discriminate between General
Relativity and its possible extensions \cite{Dimitri-rev}. For
example, some models of $f(R)$ gravity can be rejected since  do
not allow the existence of stable star configurations
\cite{Briscese, Abdalla, Bamba, Kobayashi-Maeda, Langlois,
Nojiri5}. On the other hand, the possibility of the emergence  of new
theoretical stellar structures constitute a powerful signature for any
 Extended Gravity model \cite{Laurentis, Laurentis2}. For example, one can  note that
stability of stars in modified gravity can be achieved due to the
so called {\it Chameleon Mechanism} \cite{Tsujikawa, Upadhye-Hu}.

In $f(R)$ gravity models, new scalar degrees of freedom appears.
The scalar curvature in General Relativity is defined by pressure
and density inside the star but,  in $f(R)$ gravity, scalar curvature
must be considered into dynamics as an effective new scalar field.

The structure of compact stars in perturbative $f(R)$ gravity has been recently
investigated  in some papers \cite{Arapoglu, Alavirad,
Astashenok,Goswami,Fiziev}. In this
approach, the scalar curvature $R$ is defined by the Einstein field equations
at zeroth order as a small parameter, i.e. $R\sim T$, where $T$
is the trace of the energy-momentum tensor.

In this paper, we  construct also self-consistent star models for
$f(R)=R+\alpha R^2$ gravity. We consider the case of quark stars
with very simple equation of state.  These systems could be very useful both to constrain modified gravity and
to consider stiff matter conditions in early phase transitions.

The main result
(in comparison to perturbative approach) consists on the result that one can
state that, although the gravitational mass of the star decreases with
increasing $\alpha$ (as in the perturbative approach),  outside the
star a ``gravitational sphere'' emerges. In other words, we have that the
gravitational mass of such objects (from the viewpoint of distant
observers) increases with increasing $\alpha$. This fact could constitute a new paradigm to probe such modified gravity at astrophysical scales.

The  paper is organized as  follows. In Section II, we present the
field equations for $f(R)$ gravity. For spherically symmetric
solutions of these equations, we obtain the modified
Tolman--Oppenheimer--Volkoff (TOV) equations. In Section III,
we give the description of the models in the corresponding scalar-tensor
theory. In Sec. IV, the quark star models are
presented. Discussion and conclusions  are reported
in Sec. V.

\section{Modified TOV equations in $f(R)$ gravity}

The action for $f(R)$ gravity (in units where $G=c=1$) can be
written in the form:
\begin{equation}\label{action}
S=\frac{1}{16\pi}\int d^4x \sqrt{-g}f(R) + S_{{\rm matter}},
\end{equation}
where $g$ is determinant of the metric $g_{\mu\nu}$ and $S_{\rm
matter}$ is the action of the standard perfect fluid matter.
Therefore the  Hilbert-Einstein action is  replaced by a
generic function $f(R)$ of the Ricci scalar $R$.

For solutions describing  stellar objects,  one can assume that metric is
spherically symmetric with two independent functions of radial
coordinate, that is:
\begin{equation}\label{metric}
    ds^2= -e^{2\psi}dt^2 +e^{2\lambda}dr^2 +r^2 d\Omega^2.
\end{equation}
Varying the action with respect to  $g_{\mu\nu}$ gives  the
field equations  for metric functions:
\begin{equation}\label{field}
f'(R)G_{\mu \nu }-\frac{1}{2}(f(R)-f'(R)R)g_{\mu \nu }-(\nabla
_{\mu }\nabla _{\nu }-g_{\mu \nu }\Box )f'(R)=8 \pi T_{\mu \nu }.
\end{equation}
Here $G_{\mu\nu}=R_{\mu\nu}-\frac{1}{2}Rg_{\mu\nu}$ is the
Einstein tensor, $f'(R)=df(R)/dR$ is the derivative of $f(R)$ with
respect to the scalar curvature and $T_{\mu \nu }$ is the
energy--momentum tensor. For a perfect fluid, we have
$T_{\mu\nu}=\mbox{diag}(e^{2\psi}\rho, e^{2\lambda}p, r^2p,
r^{2}\sin^{2}\theta p)$, where $\rho$ is the matter density and
$p$ is the pressure.

The components of the  field equations are nothing else but the
Tolmann-Oppenheimer-Volkov equations for $f(R)$
gravity:

\be \label{TOV1} \frac{f'(R)}{r^2}\frac{d}{dr}\left
[r\left(1-e^{-2\lambda }\right)\right]=8\pi
\rho+\frac{1}{2}\left(f'(R)R-f(R)\right)+e^{-2\lambda}\left[\left(\frac{2}{r}-\frac{d\lambda}{dr}\right)\frac{d
f'(R)}{dr}+\frac{d^{2}f'(R)}{dr^2}\right] \ee

\be \label{TOV2} \frac{f'(R)}{r}
\left[2e^{-2\lambda}\frac{d\psi}{dr}-\frac{1}{r}\left(1-e^{-2\lambda}\right)\right]
=8\pi
p+\frac{1}{2}\left(f'(R)R-f(R)\right)+e^{-2\lambda}\left(\frac{2}{r}+\frac{d\psi}{dr}\right)\frac{df'(R)}{dr}\ee
From the conservation equations  for the energy-momentum tensor, $\nabla^\mu
T_{\mu\nu}=0$, the hydrostatic condition equilibrium follows:
\begin{equation}\label{hydro}
    \frac{dp}{dr}=-(\rho
    +p)\frac{d\psi}{dr}.
\end{equation}
For $f(R)=R$ these equations reduce to the ordinary TOV equations of General Relativity. In $f(R)$
gravity, the scalar curvature is dynamical variable and the equation
for $R$ can be obtained by taking into account the  trace of field equations
(\ref{field}). We have
\begin{equation}\label{TOV3}
3\square f'(R)+f'(R)R-2f(R)=-{8\pi}(\rho-3p),\quad\mbox{where}\quad
e^{2\lambda}\square=-e^{2\lambda-2\psi}\frac{\partial^2}{\partial
t^2}+\left(\frac{2}{r}+\frac{d\psi}{dr}-\frac{d\lambda}{dr}\right)\frac{\partial}{\partial
r}+\frac{\partial ^2}{\partial r^2}\,.
\end{equation}
It is a Klein-Gordon-like equation. For $f(R)=R$ this equation is reduced to the equality

\be \label{R0} R={8\pi}(\rho-3p). \ee
Inside the star the Eqs. (\ref{TOV1}), (\ref{TOV2}),
(\ref{hydro}), (\ref{TOV3}) can be solved numerically for a given
Equation of State (EoS) $p=f({\rho})$ and initial conditions $\lambda(0)=0$,
$R(0)=R_{c}$, $R'(0)=0$ and ${\rho}(0)={\rho}_{c}$.

Outside the star, the solution is defined by the Eqs. (\ref{TOV1}),
(\ref{TOV2}), (\ref{TOV3}) where  one needs  to put $\rho=p=0$. We have
 to use the junction conditions on the surface of the star, that is
($r=r_{s}$):
$$\lambda_{in}(r_{s})=\lambda_{out}(r_{s}),\quad
R_{in}(r_{s})=R_{out}(r_{s}), \quad
R'_{in}(r_{s})=R'_{out}(r_{s}).$$
A mass parameter $m(r)$ can be defined according to the relation:
\begin{equation}
\label{mass}
    e^{-2\lambda}=1-\frac{2m}{r}.
\end{equation}
The asymptotic flatness requirement gives the constraints on the scalar
curvature and the mass parameter:
$$
\lim_{r\rightarrow\infty}R(r)=0,\quad
\lim_{r\rightarrow\infty}m(r)=\mbox{const}.
$$
For the following considerations, it is convenient to use dimensional variables $m\rightarrow m
M_{\odot}$ and $r\rightarrow r_{g} r$ where
$r_{g}=GM_{\odot}/c^{2}$. This assumption allows to refer to typical stellar structures.

\section{The scalar-tensor  gravity picture}

Let us consider the analogue modified gravity  description  in terms of
scalar-tensor gravity (see \cite{Kokkotas, Kokkotas-1}). For $f(R)$
gravity one can construct the equivalent Brans-Dicke-like  theory where the
action for the gravitational sector is:
 \be S_{g}=\frac{1}{16\pi}\int
d^{4} x \sqrt{-g}\left(\Phi R - U(\Phi)\right). \ee
Here $\Phi=df(R)/dR$ is gravitational scalar and
$U(\Phi)=Rf'(R)-f(R)$ is potential. It is worth noticing that standard Brans-Dicke shows a kinetic term instead of a potential like in this case.

Using the  conformal transformation
$\tilde{g}_{\mu\nu}=\Phi g_{\mu\nu}$, one can write the action in the
Einstein frame as
 \be S_{g}=\frac{1}{16\pi G}\int d^{4} x
\sqrt{-\tilde{g}}\left(\tilde{R}
-2\tilde{g}^{\mu\nu}\partial_{\mu}\phi\partial_{\nu}\phi-2V(\phi)\right),
\ee where $\phi=\sqrt{3}\ln\Phi/2$. In such a frame,  the potential   becomes
$V(\phi)=\Phi^{-2}(\phi)U(\Phi(\phi))/2$.
It is convenient to choose the space-time metric in a  form that formally
coincided with (\ref{metric}). We can redefine the  functions $\psi$ and
$\lambda$ as

\be \label{metric2} d\tilde{s}^{2}=\Phi
ds^{2}=-e^{2\tilde{\psi}}dt^{2}+e^{2\tilde{\lambda}}\tilde{dr}^{2}+\tilde{r}^2d\Omega^2.
\ee
In Eq. (\ref{metric2}) $\tilde{r}^2=\Phi r^{2}$,
$e^{2\tilde{\psi}}=\Phi e^{2\psi}$ and from equality
$$
\Phi e^{2\lambda}dr^{2}=e^{2\tilde{\lambda}}d\tilde{r}^{2}
$$
it follows that
$$
e^{-2\lambda}=e^{-2\tilde{\lambda}}\left(1-\tilde{r}\phi'(\tilde{r})/\sqrt{3}\right)^{2}.
$$
Therefore the mass parameter $m(\tilde{r})$ can be obtained from
$\tilde{m}(\tilde{r})$ as

\be m(\tilde{r})=
\frac{\tilde{r}}{2}\left(1-\left(1-\frac{2\tilde{m}}{\tilde{r}}\right)\left(1-\tilde{r}\phi'(\tilde{r})/\sqrt{3}\right)^{2}\right)e^{-\phi/\sqrt{3}}.
\ee
The resulting equations for metric functions $\tilde{\lambda}$ and
$\tilde{\psi}$ (tildes in the following  will be  omitted for simplicity):

\be \label{TOV1-1} \frac{1}{r^2}\frac{d}{dr}\left
[r\left(1-e^{-2{\lambda} }\right)\right]=8\pi
e^{-4\phi/\sqrt{3}}\rho+e^{-2{\lambda}}\left(\frac{d\phi}{dr}\right)^{2}+V(\phi),
\ee

\be \label{TOV2-1} \frac{1}{r}
\left[2e^{-2{\lambda}}\frac{d{\psi}}{dr}-\frac{1}{r}\left(1-e^{-2{\lambda}}\right)\right]
=8\pi
e^{-4\phi/\sqrt{3}}p+e^{-2{\lambda}}\left(\frac{d\phi}{dr}\right)^{2}-V(\phi).\ee
The hydrostatic equilibrium condition can be  rewritten as
\begin{equation}\label{hydro-1}
    \frac{dp}{dr}=-(\rho
    +p)\left(\frac{d{\psi}}{dr}-\frac{1}{\sqrt{3}}\frac{d\phi}{dr}\right).
\end{equation}
Finally the last field equation  for the scalar field is equivalent
to Eq. (\ref{TOV3}) in $f(R)$ theory:

\be \label{TOV3-1} \square
\phi-\frac{1}{2}\frac{dV(\phi)}{d\phi}=-\frac{4\pi}{\sqrt{3}}
e^{-4\phi/\sqrt{3}}(\rho-3p). \ee
The first two equations in fact coincide with the ordinary TOV
equations with redefined matter density and pressure where also the
energy density and the pressure of the scalar field $\phi$ are included.
It is worth  noticing  that the potential $V(\phi)$ can be written in explicit
form only for simple $f(R)$ models. For example for $f(R)=R+\alpha
R^2$,  one  obtains that

\be V(\phi)=\frac{1}{8\alpha}\left(1-e^{-2\phi/\sqrt{3}}\right)^2.
\ee
Considering simple power models like  $f(R)=R+\alpha R^{m}$,  we have

\be V(\Phi)=D\Phi^{-2}\left(\Phi-1\right)^{\frac{m}{m-1}}, \quad\mbox{with}\quad
D=\frac{m-1}{2m^{\frac{m}{m-1}}}\alpha^{\frac{1}{1-m}}, \quad
\Phi=e^{2\phi/\sqrt{3}}. \ee
These considerations can be applied to quark star models by assuming suitable EoS.

\section{Quark star models in $f(R)=R+\alpha R^{2}$ gravity}

\begin{figure}
  \includegraphics[scale=0.7]{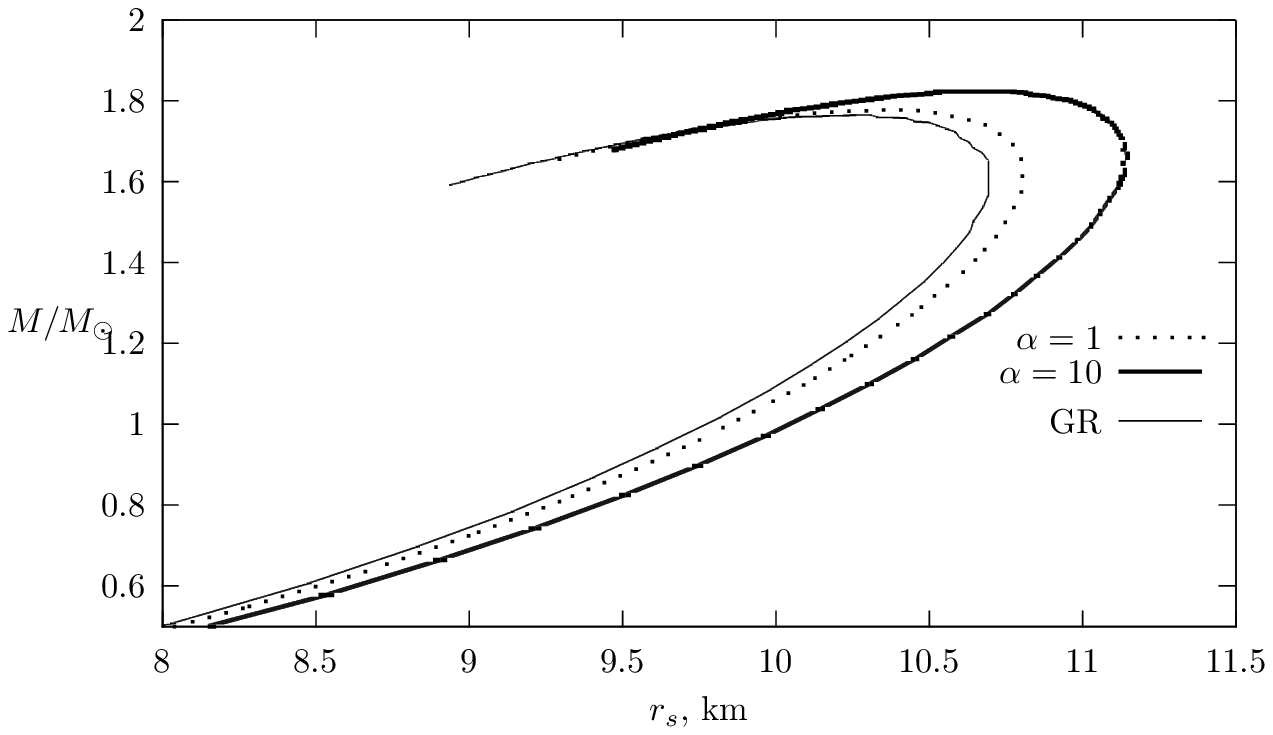} \includegraphics[scale=0.7]{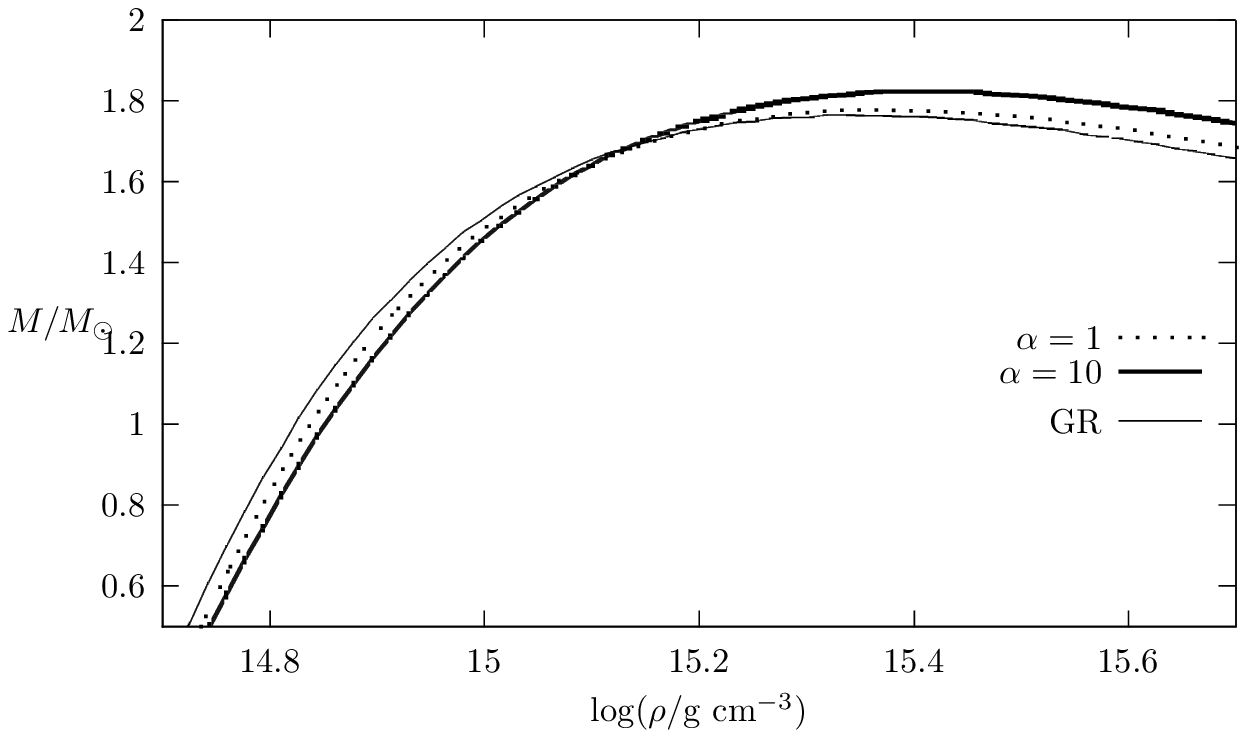}\\
  \caption{The mass-radius (upper panel) and mass-central density (lower panel) diagram in model $f(R)=R+\alpha R^2$ and in GR for quark stars with $B=60$ MeV/fm$^{3}$ and $c=0.28$.}
\end{figure}

\begin{figure}
  \includegraphics[scale=0.7]{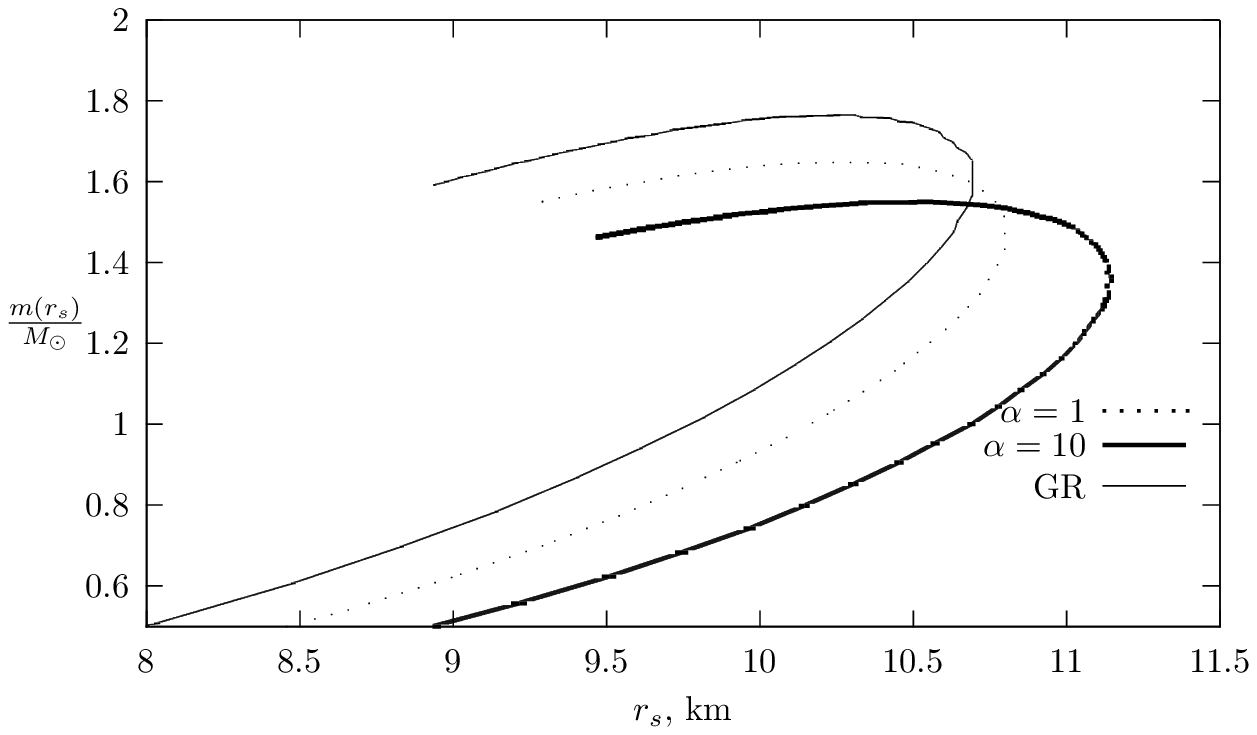} \includegraphics[scale=0.7]{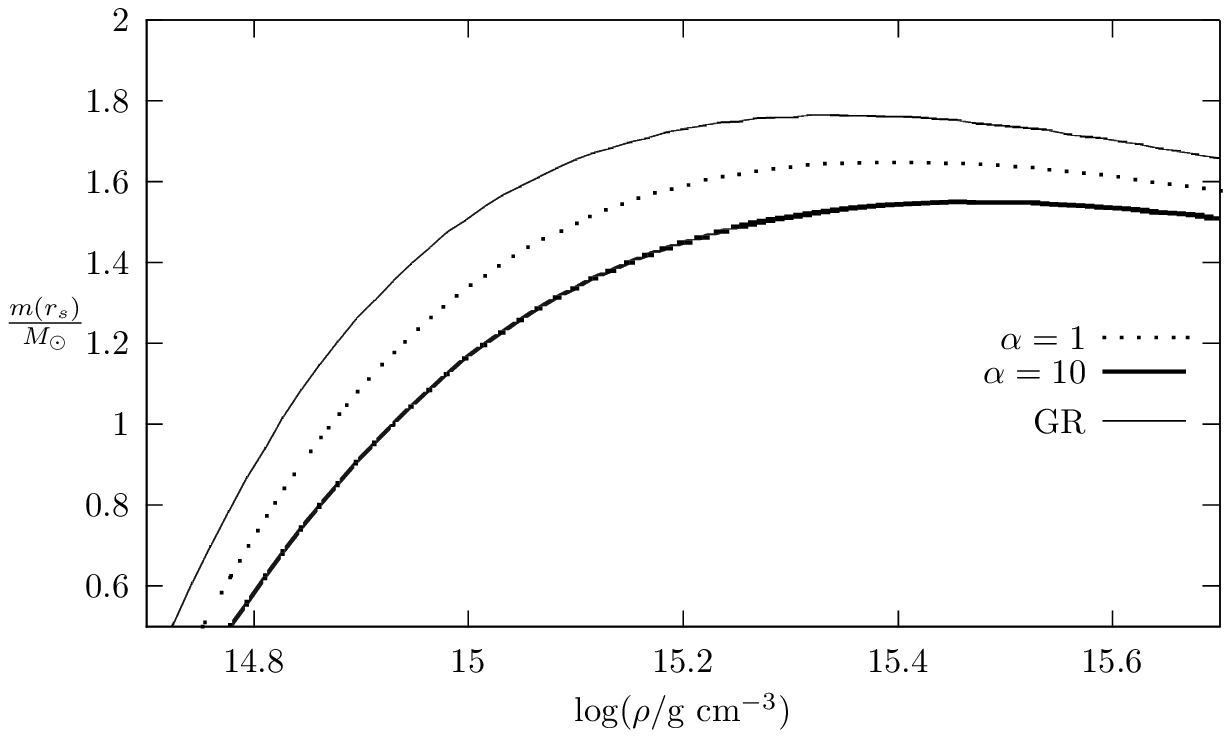}\\
  \caption{The $m(r_{s})$-radius diagram (upper panel) and $m(r_{s})$-central density (lower panel) diagram in model $f(R)=R+\alpha R^2$ for quark stars with $B=60$ MeV/fm$^{3}$ and $c=0.28$.}
\end{figure}

\begin{figure}
  \includegraphics[scale=1.1]{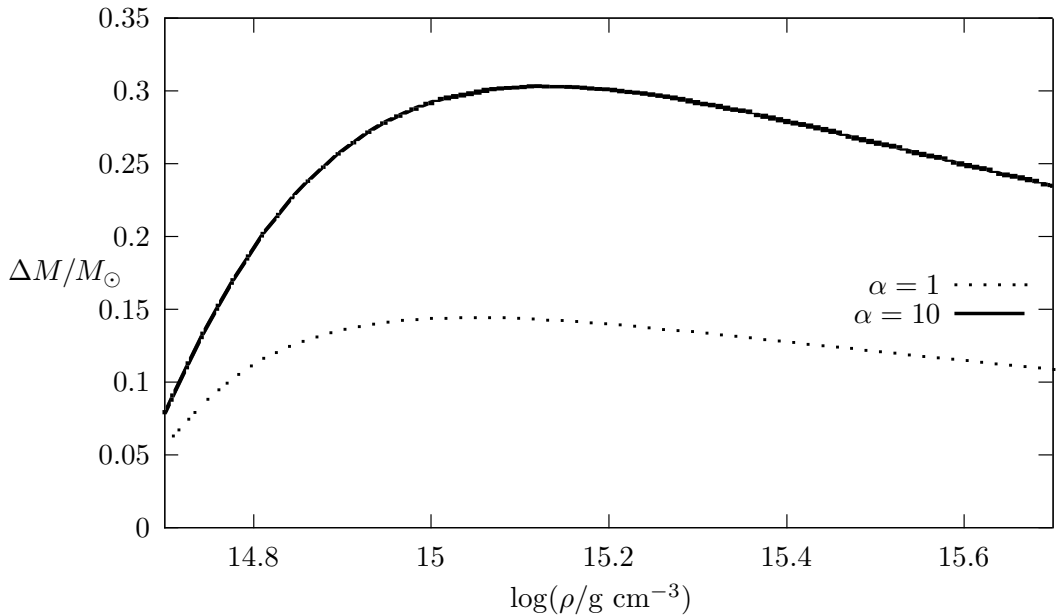}\\
  \caption{The dependence $\Delta M=M-m(r_{s})$ from central density in model $f(R)=R+\alpha R^2$ for quark stars with $B=60$ MeV/fm$^{3}$ and $c=0.28$.}
\end{figure}

\begin{figure}
  \includegraphics[scale=0.7]{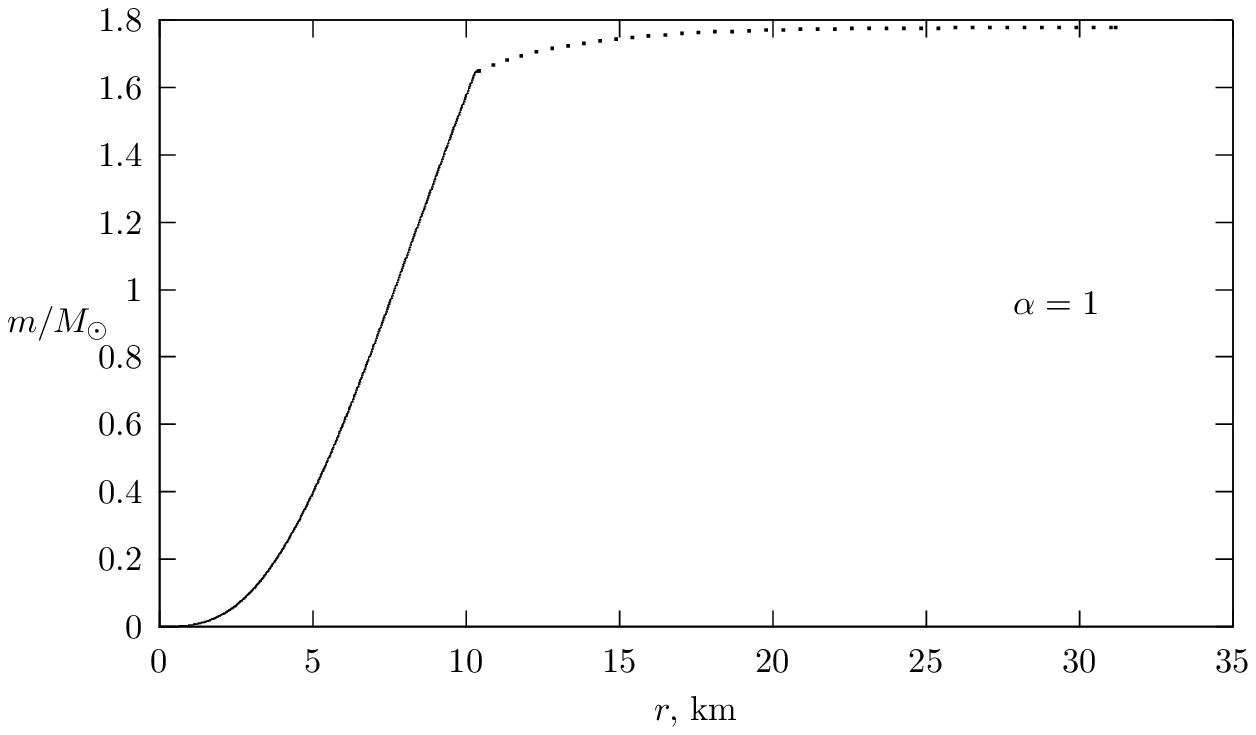} \includegraphics[scale=0.7]{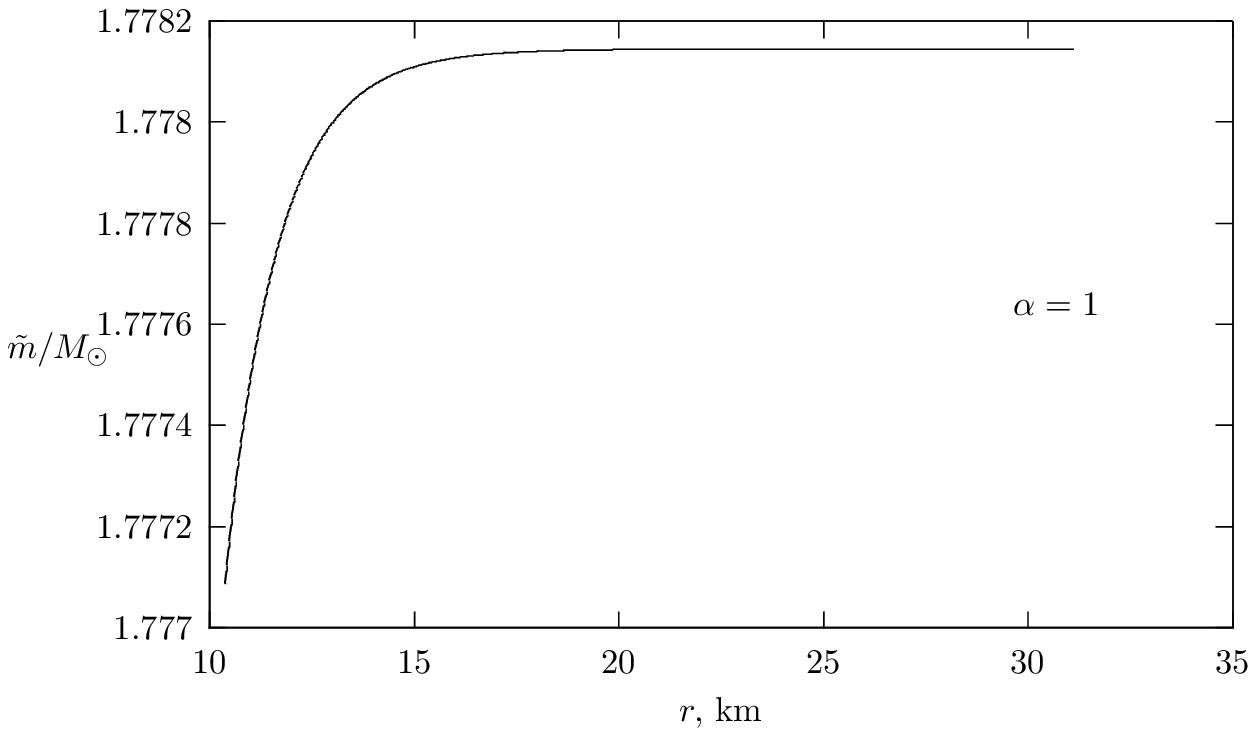}\\
  \includegraphics[scale=0.7]{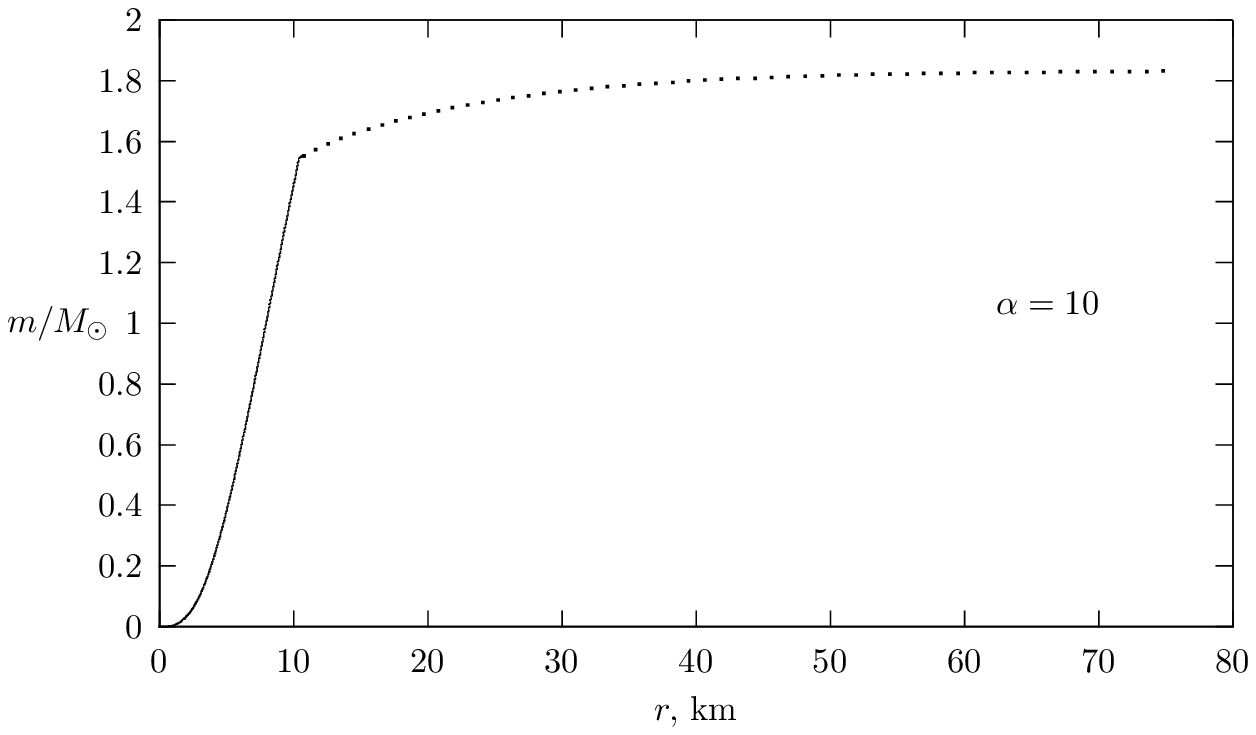} \includegraphics[scale=0.7]{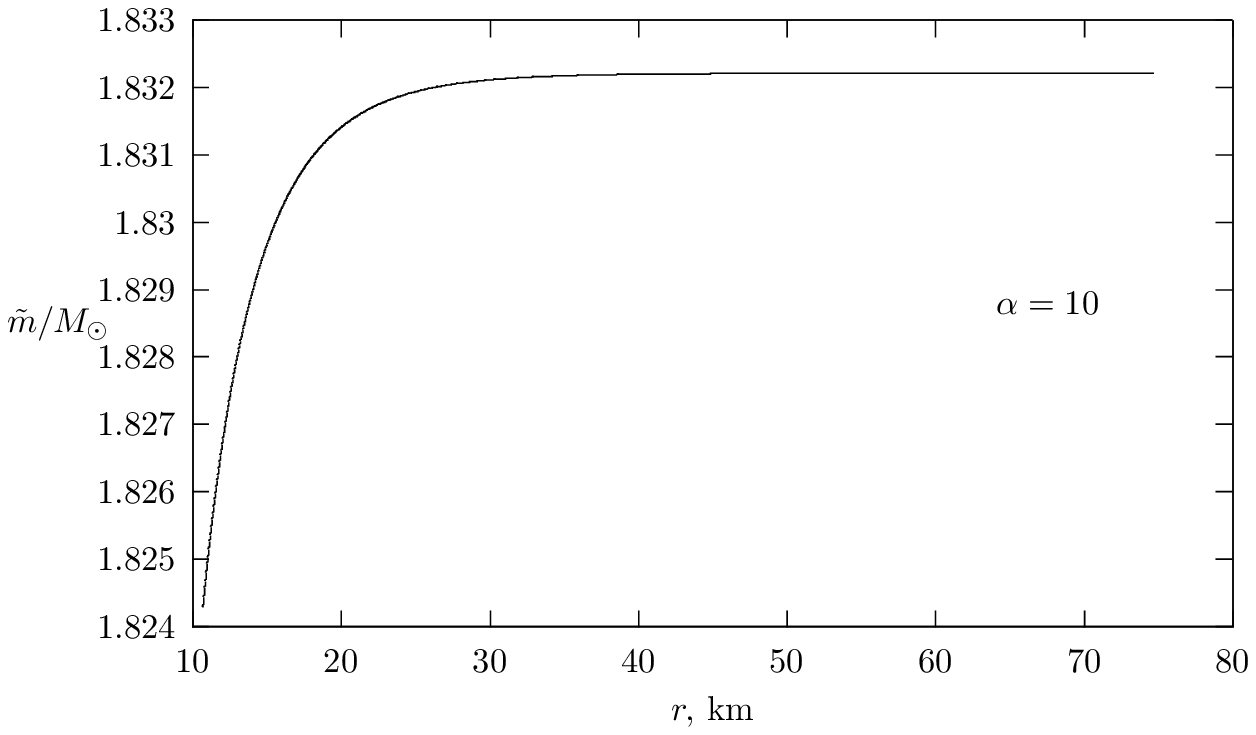}\\
  \caption{The mass parameter $m(r)$ profile inside (bold curves) and outside the star (dotted curves) (left panel)
  in model $f(R)=R+\alpha R^2$ for stars configuration with maximal mass ($B=60$ MeV/fm$^{3}$ and $c=0.28$). For comparison we give the dependence
  $\tilde{m}(r)$ outside the star in corresponding scalar-tensor theory (right panel).}
\end{figure}

\begin{figure}
  \includegraphics[scale=0.7]{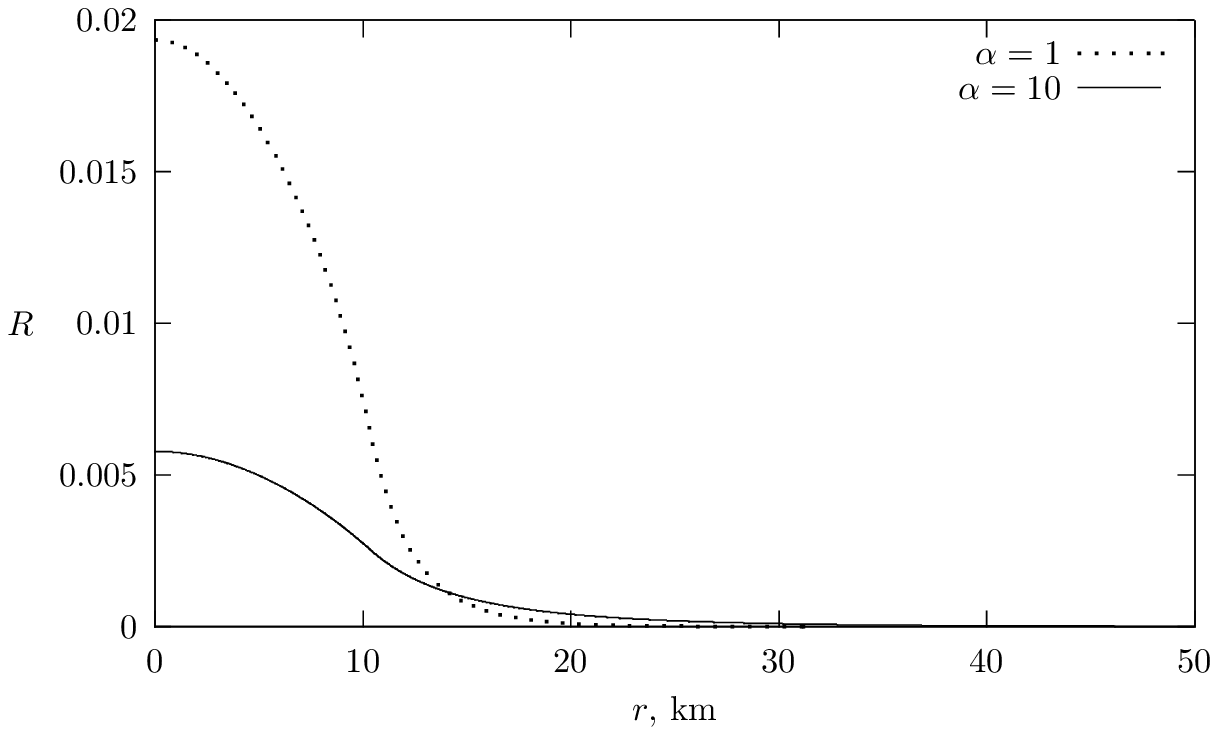}\includegraphics[scale=0.7]{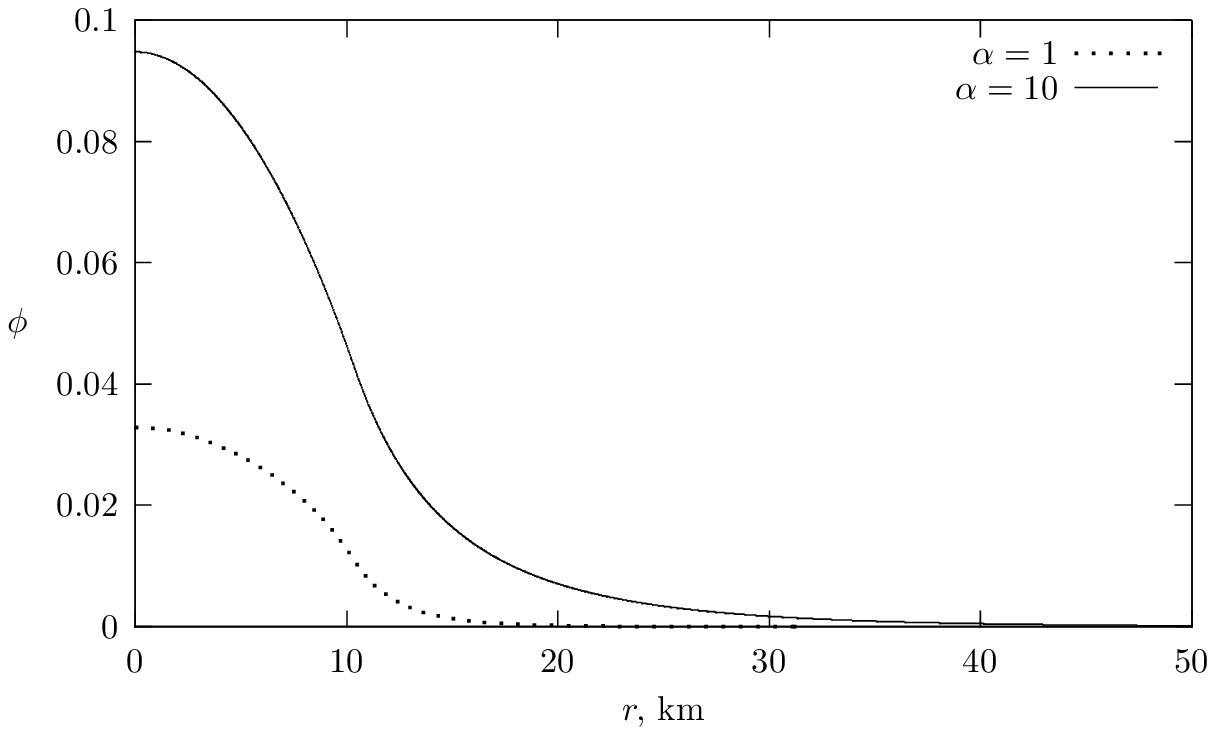}\\
  \caption{The scalar curvature $R(r)$ (left panel)
  in model $f(R)=R+\alpha R^2$ for stars configuration with maximal mass ($B=60$ MeV/fm$^{3}$ and $c=0.28$). On right panel we depict the corresponding profile of
  scalar field $\phi$ in corresponding scalar-tensor theory.}
\end{figure}

A quark star is a self-gravitating system consisting of deconfined $u$, $d$ and
$s$ quarks and electrons \cite{Witten}. Deconfined quarks form
colour superconductor system, leading to a softer equation of state in
comparison with the standard hadron matter.

In the frame of the so called  {\it MIT bag model} one can obtain a simple equation of
state for quark matter: \be \label{QEOS} p=c(\rho-4B),\ee where
$B$ is the bag constant. The value of parameter $c$ depends on the
chosen mass of strange quark $m_s$ and QCD coupling constant. For
$m_s=0$, the parameter is $c=1/3$ as for radiation. For more realistic model with
$m_{s}=250$ MeV, we have $c=0.28$. The value of $B$ lies in the
interval $0.98<B<1.52$ in units of $B_{0}=60$ MeV/fm$^{3}$
\cite{Sterg}.

The solution of Eqs. (\ref{TOV1})-(\ref{TOV3}) can be achieved by
using a perturbative approach (see \cite{Arapoglu,Alavirad} for
details). In this perturbative approach,  the
curvature scalar  cannot considered as an additional degree of freedom since its
value is fixed by the  relation (\ref{R0}). Perturbative calculations
show that gravitational mass of star decreases with increasing
$\alpha$ for the  $f(R)=R+\alpha R^2$ model.

Let us  consider now the system (\ref{TOV1})-(\ref{TOV3}) for
$f(R)=R+\alpha R^2$ model assuming that $R$ is an independent
function. One has to note that authors of \cite{Naf} estimated the upper
limit for $\alpha$ as $\sim 5\times 10^{15}$ cm$^2$$=2.3\times
10^5 r_{g}^{2}$  from binary pulsar data.

We find that for each value of central density, we have the solution
with required asymptotic $R\rightarrow 0$ at $r\rightarrow\infty$
only for a unique value of $R(0)$. In the scalar-tensor description,  this
uniqueness of $R(0)$ is equivalent to a fine-tuning conditions for the scalar field
$\phi$ at the center of star for a given density.

It is interesting to note that the gravitational mass of star $M$
(by calculating as asymptotic mass value   the parameter $m(r)$)
increases with increasing $\alpha$. At first glance this
contradiction (in comparison with results coming from perturbative
approach) shows that perturbative approach is inadequate to deal with such problems in
$f(R)$ gravity. However a detailed investigation leads to the conclusion
that perturbative approach is neither inadequate nor incomplete. One
can say that the  increasing of mass occurs on the ``gravitational sphere''
outside the star as some ``effective mass''. Without this
``sphere'' the gravitational mass of star decreases (of course we
cannot actually distinguish  the star from this sphere but this
interpretation has a right to exist). In the framework  of the perturbative
approach,  one cannot account for  the existence of such a sphere because
the Schwarzschild solution outside the star is assumed.

In the conformal frame of the corresponding scalar-tensor theory, we
 have also the  so called dilaton sphere (``disphere'') outside the star
but its contribution to gravitational mass for distant observer
is negligible.

The mass-radius and the mass-central density diagram for quark stars
with realistic EOS ($c=0.28$, $B=60$ MeV/fm$^3$) are represented in
Fig. 1. The radius of star increases in comparison with General
Relativity.  The star configurations with maximal mass correspond
to larger central densities (see Table I). We consider also
an EoS with $c=0.28$ and $B=60$ MeV/fm$^3$ (Table 2).
In Fig. 2 we plot the dependence of mass parameter $m(r_{s})$ (the
value of mass parameter on the surface of star) against the radius and the
central density. In Fig. 3 the dependence $\Delta M=M-m(r_{s})$
from central density is presented.

The mass parameter profile $m(r)$ for star configurations with
maximal mass is represented  in Fig. 4. One can see that the radius of
gravitational sphere increases with growing $\alpha$. The mass
parameter $\tilde{m}$ reaches the value close to the maximal on the
surface of the star. In Fig. 5 the scalar curvature and the scalar field
(in terms of scalar-tensor description) as functions of radial
coordinate are given.

One has to note that the initial condition for $R$ in the center correlates with the
value $R^{(0)}=8\pi(\rho-3p)$, i.e. the scalar curvature in General
Relativity. For $\alpha\rightarrow 0$, the  scalar curvature
$R(0)\rightarrow R^{(0)}$ as it is  expected.

\begin{table}
\label{Table1}
\begin{centering}
\begin{tabular}{|c|c|c|c|c|c|}
  \hline

$\alpha,$ & $M_{max}$, & $m(r_{s}),$ & $r_{s}$,  & $\rho_c$, & $R_{c}$, \\
 $r_{g}^{2}$ &  $M_{\odot}$ & $M_{\odot}$ & km & 10$^{15}$ g/cm$^{3}$ & $10^{-3} r_{g}^{-2}$ \\
  \hline
  0 & 1.764 & 1.764 & 10.26 & 2.17 & 28.62  \\
  1 & 1.778 & 1.649 & 10.38 & 2.26 & 19.34  \\
  10 & 1.832 & 1.552 & 10.68 & 2.54 & 5.78 \\
  \hline
\end{tabular}
\caption{Quark star properties using the simple model (\ref{QEOS})
with $c=0.28$ and $B=60$ MeV/fm$^{3}$ for $f(R)=R+\alpha R^2$
gravity.}
\end{centering}
\end{table}

\begin{table}
\label{Table1}
\begin{centering}
\begin{tabular}{|c|c|c|c|c|c|}
  \hline

$\alpha,$ & $M_{max}$, & $m(r_{s}),$ & $r_{s}$,  & $\rho_c$, & $R_{c}$, \\
 $r_{g}^{2}$ &  $M_{\odot}$ & $M_{\odot}$ & km & 10$^{15}$ g/cm$^{3}$ & $10^{-3} r_{g}^{-2}$ \\
  \hline
  0 & 1.883 & 1.883 & 10.54 & 2.09 & 22.01  \\
  1 & 1.901 & 1.772 & 10.61 & 2.26 & 19.34  \\
  10 & 1.966 & 1.681 & 10.92 & 2.54 & 5.38 \\
  \hline
\end{tabular}
\caption{Quark star properties using the simple model (\ref{QEOS})
with $c=0.31$ and $B=60$ MeV/fm$^{3}$ for $f(R)=R+\alpha R^2$
gravity.}
\end{centering}
\end{table}

One can see that the deviation of the  mass-radius relation from General
Relativity, in principle, is not so large. Taking into account that
there are no precise radius measurements for  stars, one cannot
hope that the mass-radius relation gives argument in favor (or not) of
modified gravity.

However,  in principle,  it is possible to discriminate modified
theories of gravity from General Relativity due to the redshift of
the surface atomic lines. The gravitational redshift $z$  of
thermal spectrum detected at infinity can be calculated as

\be z=\ e^{-\psi}-1.\ee
In General Relativity we have simply
$$
z(r_{s})=\frac{1}{\sqrt{1-2M/r_{s}}}-1.
$$
In the case of modified gravity, we have another dependence on the
surface redshift from gravitational mass. Calculations give the
following results. In General relativity for maximal mass we have
$z(r_{s})=0.424$ ($c=0.28$ and $B=60$ MeV/fm$^{3}$). In the case
of quadratic gravity with the surface redshift for star with
maximal mass, it  is 0.431 ($\alpha=1$) and 0.458 ($\alpha=10$).

 Of course the measurement of $z(r_{s})$ can constrain the 
 theories of gravity only in  the case where the mass of the star is
measured with high precision (for example by precise measurements by binary systems dynamics).
Another requirement for discriminating between General Relativity
and modified gravity is that one should  know the realistic
equation of state in extreme details. This information is not available with today facilities but could be acquired by forthcoming experiments like the {\it Large Observatory For X-ray Timing} (LOFT) whose one of the main scientific goals is to select reliable equations of state for compact objects in strong gravity regimes \cite{LOFT}.

Another difficulty is that present accuracy on redshift
measurements is not sufficient yet for constraining gravity in
strong regime. In future, one can hope that the increasing number
of good quality data on the thermal emission with mass
measurements could help to distinguish General Relativity from
models of $f(R)$ gravity. Again LOFT experiment could be useful for this task.

In this framework, the question about possible instabilities
during the star formation arises. First of all, we have to note
that the scalar curvature inside the matter sphere is smaller in
comparison with General Relativity for $\alpha>0$ for the
considered models. It is also known that adding the higher
derivative term ($\sim R^{\beta}$, $1<\beta\leqslant 2$) to the
standard Hilbert-Einstein action could cure the singularity (for
details see \cite{Bamb}).

For  the quadratic gravity, it is easy to see this fact by using
the scalar-tensor description.  Eq. (\ref{TOV3-1}) for the scalar field
can be rewritten as:

\be \square \phi -\frac{dV_{eff}}{d\phi}=0, \quad
V_{eff}=\frac{1}{2}V+\pi e^{-4\phi/\sqrt{3}}(\rho-3p).\ee
The effective squared mass of the scalar field is defined as \be
m^{2}_{eff}=\frac{d^{2}V_{eff}}{d\phi^{2}}=\frac{1}{2}\frac{d^{2}V}{d\phi^{2}}+\frac{16\pi}{3}e^{-4\phi/\sqrt{3}}(\rho-3p).
\ee

The first term for the model with $R$-squared term is positive.
For quark stars, the second term is also positive (because
$p<\rho/3$). Therefore effective mass has a real value. It is
well-known that in this case the solution corresponds to the
minimum of potential (this minimum corresponds to some value of
curvature). For radial modes of the perturbations, we have
decaying solutions (see \cite{Langlois}). Therefore the considered
gravity model (with the quark equation of state) could become free
of curvature singularity.

\section{Conclusion}

We have considered  realistic quark star models in nonperturbative $f(R)$
gravity and obtained the  parameters of stars in $f(R)=R+\alpha R^2$
model. The key issue of such a  non-perturbative approach is that one needs
to consider the scalar curvature as an  independent function. The
shooting method of solution gives that there is  a unique value of the
curvature at the center of the star where  solution has the required
asymptotic behavior. This fine-tuning for $R$ at the center of the star is
equivalent to the fine-tuning of the scalar field $\phi$ in the corresponding
scalar-tensor theory. For a distant observer,  the gravitational mass
of the star increases with increasing $\alpha$ ($\alpha>0$). One can
say that the increasing of the mass occurs on the ``gravitational sphere''
outside the star with some ``effective mass''. In the corresponding conformally transformed scalar-tensor
theory,  we  have also the dilaton sphere (or the
disphere) outside the star but its contribution to the gravitational
mass for distant observer is negligible.

The considered approach can be applied for the analysis of structure
of neutron stars in modified gravity. The calculations show that
for realistic hyperon EoS we have, in principle, the same effects as
for quark stars.

Although the deviation of mass-radius relation from General
Relativity is sufficiently small, it is possible to discriminate
modified theory of gravity from General Relativity due to the redshift
of the surface atomic lines. In $f(R)=R+\alpha R^2$ gravity the
surface redshift grows with the  increasing of the parameter $\alpha$.

\acknowledgments

This work is supported in part by project 14-02-31100 (RFBR,
Russia) and by project 2058 (MES, Russia) (AVA). SC is supported by INFN ({\it iniziative
specifiche} TEONGRAV and QGSKY).


\begin{thebibliography}{99}

\bibitem{Perlmutter} S.~Perlmutter {\it et al.}  [Supernova Cosmology Project Collaboration], Astrophys.\ J.\  {\bf 517}, 565 (1999) arXiv:astro-ph/9812133.

\bibitem{Riess1} A.~G.~Riess {\it et al.}  [Supernova Search Team Collaboration], Astron.\ J.\  {\bf 116}, 1009 (1998) arXiv:astro-ph/9805201.

\bibitem{Riess2} A.~G.~Riess {\it et al.}  [Supernova Search Team Collaboration], Astrophys.\ J.\  {\bf 607}, 665 (2004) arXiv:astro-ph/0402512.

\bibitem{Spergel} D.~N.~Spergel {\it et al.}  [WMAP Collaboration], Astrophys.\ J.\ Suppl.\  {\bf 148}, 175 (2003) arXiv:astro-ph/0302209.

\bibitem{Schmidt} C. Schimdt et al., Astron. Astrophys. {\bf 463}, 405 (2007).

\bibitem{McDonald} P. McDonald et al., Astrophys. J. Suppl. {\bf 163}, 80 (2006).

\bibitem{Capozziello1} S. Capozziello, Int. J. Mod. Phys.  D {\bf  11}, 483 (2002).

\bibitem{Capozziello2} S. Capozziello, S. Carloni, A. Troisi, Recent Res. Dev. Astron. Astrophys. {\bf 1}, 625 (2003).

\bibitem{Odintsov1} S. Nojiri, S.D. Odintsov, Phys. Rev. D {\bf 68}, 123512 (2003); Phys. Lett. B  {\bf 576}, 5 (2003).

\bibitem{Turner} S.~M.~Carroll, V.~Duvvuri, M.~Trodden and M.~S.~Turner, Phys. Rev. D. {\bf 70},  043528 (2004).

\bibitem{Olmo}   G. J. Olmo,  Int.J.Mod.Phys. D {\bf 20}, 413 (2011).

\bibitem{Odintsov-3} S. Nojiri and S. D. Odintsov, Phys. Rept. {\bf 505}, 59 (2011) arXiv:1011.0544 [gr-qc]; eConf C 0602061, 06 (2006) [Int. J.
Geom. Meth. Mod. Phys. {\bf 4}, 115 (2007)] [hep-th/0601213]; arXiv:1306.4426 [gr-qc].

\bibitem{Capozziello_book} S. Capozziello and V. Faraoni, \textit{Beyond Einstein Gravity} (Springer) New York  (2010).

\bibitem{Capozziello4} S. Capozziello and M. De Laurentis, Phys. Rept.{\bf  509}, 167 (2011) arXiv:1108.6266[gr-qc].

\bibitem{Cruz} A. de la Cruz-Dombriz and D. Saez-Gomez, Entropy {\bf 14}, 1717 (2012) arXiv:1207.2663 [gr-qc].

\bibitem{Weinberg} S. Weinberg, Rev. Mod. Phys. {\bf 61}, 1 (1989).

\bibitem{Bancall} N.A. Bahcall et al., Science {\bf 284}, 1481
(1999);

K. Bamba, S. Capozziello, S. Nojiri, S.D. Odintsov, Astrophys.
Space Sci. 342, 155 (2012);

A. Joyce, B. Jain, J. Khoury, M. Trodden, arXiv:1407.0059
[astro-ph.CO].

\bibitem{Demianski} M. Demianski et al., Astron. Astrophys. {\bf 454}, 55 (2006).

\bibitem{Perrotta} V. Perrotta, C. Baccagalupi, S. Matarrese, Phys. Rev. D {\bf  61}, 023507 (2000).

\bibitem{Hwang} J.C. Hwang, H. Noh, Phys. Lett.  B {\bf 506}, 13 (2001).

\bibitem{Dimitri-rev} D.~Psaltis, Living Reviews in Relativity, {\bf 11}, 9 (2008) arXiv:0806.1531 [astro-ph].

\bibitem{Briscese} F.~Briscese, E.~Elizalde, S.~Nojiri and S.D.~Odintsov, Phys.\ Lett.\ B {\bf  646}, 105 (2007) arXiv:hep-th/0612220.

\bibitem{Abdalla} M.C.B. Abdalla, S.~Nojiri and S.D.~Odintsov, Class.\ Quant.\ Grav.\  {\bf 22}, L35 (2005)
arXiv:hep-th/0409177.

\bibitem{Bamba} K.~Bamba, S.~Nojiri and S.~D.~Odintsov, JCAP {\bf 0810}, 045 (2008) arXiv:0807.2575 [hep-th]; Phys. Lett. B {\bf 698}, 451 (2011) arXiv:1101.2820[gr-qc].

\bibitem{Kobayashi-Maeda} T.~Kobayashi and K.~i.~Maeda, Phys.\ Rev.\ D {\bf  78}, 064019 (2008) arXiv:0807.2503
[astro-ph].

\bibitem{Langlois} E. Babichev, D. Langlois, Phys. Rev. D 81, 124051
(2010) arXiv:0911.1297 [gr-qc].

\bibitem{Nojiri5} S.~Nojiri and S.~D.~Odintsov, Phys.\ Lett.\   B {\bf  676}, 94 (2009) arXiv:0903.5231 [hep-th].

\bibitem{Laurentis} S.~Capozziello, M.~De Laurentis, I.~De Martino, M.~Formisano and S.~D.~Odintsov, Phys.\ Rev.\ D {\bf 85}, 044022 (2012) arXiv:1112.0761 [gr-qc].

\bibitem{Laurentis2} S.~Capozziello, M.~De Laurentis, S.D.~Odintsov and A.~Stabile, Phys.\ Rev.\ D {\bf 83}, 064004 (2011) arXiv:1101.0219 [gr-qc].

\bibitem{Tsujikawa} J.~Khoury and A.~Weltman, Phys. Rev. D 69, 044026 (2004)
astro-ph/0309411;

J. Khoury and A. Weltman, Phys. Rev. Lett. 93, 171104 (2004)
arXiv:astro-ph/0309300.

\bibitem{Upadhye-Hu} A.~Upadhye and W.~Hu, Phys.\ Rev.\ D {\bf  80}, 064002 (2009) arXiv:0905.4055 [astro-ph.CO].

\bibitem{Arapoglu} S. Arapoglu, C. Deliduman, K. Yavuz Eksi, JCAP {\bf 1107}, 020 (2011) arXiv:1003.3179v3 [gr-qc].

\bibitem{Alavirad} H. Alavirad, J.M. Weller, arXiv:1307.7977v1[gr-qc].

\bibitem{Astashenok} A. Astashenok, S. Capozziello, S. Odintsov, JCAP \textbf{12}, 040 (2013) arXiv:1309.1978 [gr-qc].\\
 A. Astashenok, S. Capozziello, S. Odintsov,Phys. Rev. D {\bf 89}, 103509 (2014)  arXiv:1401.4546 [gr-qc].\\
 A. Astashenok, S. Capozziello, S. Odintsov,  Astrophys. Space Sci. {\bf 355}, 2182  (2014).\ arXiv:1405.6663 [gr-qc] .\\
 A. Astashenok, S. Capozziello, S. Odintsov,  JCAP {\bf 01} (2015) 001,   arXiv:1408.3856 [gr-qc].
 
\bibitem{Goswami}
A. Ganguly, R. Gannouji, R. Goswami, and S. Ray Phys. Rev. D {\bf 89}, 064019 (2014).\\
R. Goswami, A. M. Nzioki, S. D. Maharaj, and S. G. Ghosh Phys. Rev. D {\bf 90}, 084011 (2014). 

\bibitem{Fiziev} P. Fiziev,  Phys. Rev. D {\bf 87}, 044053 (2013).\\
 P. Fiziev,  arXiv:1402.2813v1 [gr-qc]  (2014).\\
P.  Fiziev,  PoS (FFP14) 080,  arXiv:1411.0242v1 [gr-qc]    (2014).\\
3. P. Fiziev, K. Marinov,  arXiv:1412.3015v1 [gr-qc]  (2014).

\bibitem{Kokkotas} K.V. Staykov, D.D. Doneva, S.S. Yazadjiev, K.D.
Kokkotas, JCAP 1406, 003 (2014).

\bibitem{Kokkotas-1} K.V. Staykov, D.D. Doneva, S.S. Yazadjiev, K.D.
Kokkotas, arXiv:1407.2180 [gr-qc].

\bibitem{Witten} E.Witten, Phys. Rev. D 30, 272 (1984).

\bibitem{Sterg} N. Stergioulas, Living Rev. Rel. 6, 3 (2003) arXiv:gr-qc/0302034 [gr-qc].

\bibitem{Naf} J. Naf and P. Jetzer, Phys. Rev. D 81, 104003 (2010) arXiv:1004.2014 [gr-qc].

\bibitem{Bamb} K. Bamba, S. Nojiri, S.D. Odintsov, Phys. Lett. B 698, 451 (2011) arXiv:1101.2820 [gr-qc].

\bibitem{LOFT} http://www.isdc.unige.ch/loft/index.php/science
 \end{thebibliography}
\end{document}